\documentclass{article}

    \PassOptionsToPackage{numbers, compress}{natbib}

\usepackage[preprint]{neurips_2023}

\usepackage[utf8]{inputenc} 
\usepackage[T1]{fontenc}    
\usepackage{hyperref}       
\usepackage{url}            
\usepackage{booktabs}       
\usepackage{amsfonts}       
\usepackage{nicefrac}       
\usepackage{microtype}      
\usepackage{xcolor}         
\usepackage{enumitem}
\usepackage{amsmath}
\usepackage{graphicx} 
\usepackage{float} 
\usepackage{subfigure} 
\usepackage{multirow}
\usepackage{multicol}
\usepackage{authblk}

\title{Mapping Biological Neuron Dynamics into an Interpretable Two-layer Artificial Neural Network}

\author[a,b]{Jingyang Ma}
\author[a,b,c,1]{Songting Li }
\author[a,b,c,d,1]{Douglas Zhou}
\affil[a]{School of Mathematical Sciences, Shanghai Jiao Tong University, Shanghai 200240, China}
\affil[b]{Institute of Natural Sciences, Shanghai Jiao Tong University, Shanghai 200240, China}
\affil[c]{Ministry of Education Key Laboratory of Scientific and Engineering Computing, Shanghai Jiao Tong University, Shanghai 200240,
China}
\affil[d]{Shanghai Frontier Science Center of Modern Analysis, Shanghai Jiao Tong University, Shanghai 200240, China}

\begin{document}

\maketitle

\renewcommand{\thefootnote}{\fnsymbol{footnote}} 
\footnotetext{$^1$To whom correspondence may be addressed. Email: songting@sjtu.edu.cn, or zdz@sjtu.edu.cn.} 

\begin{abstract}
  Dendrites are crucial structures for computation of an individual neuron. It has been shown that the dynamics of a biological neuron with dendrites can be approximated by artificial neural networks (ANN) with deep structure. However, it remains unclear whether a neuron can be further captured by a simple, biologically plausible ANN. In this work, we develop a two-layer ANN, named as dendritic bilinear neural network (DBNN), to accurately predict both the sub-threshold voltage and spike time at the soma of biological neuron models with dendritic structure. Our DBNN is found to be interpretable and well captures the dendritic integration process of biological neurons including a bilinear rule revealed in previous works. In addition, we show DBNN is capable of performing diverse tasks including direction selectivity, coincidence detection, and image classification. Our work proposes a biologically interpretable ANN that characterizes the computation of biological neurons, which can be potentially implemented in the deep learning framework to improve computational ability.
\end{abstract}

\section{Introduction}

Neurons are fundamental units of the nervous system to perform complex computational functions. Dendrites, which are the branched extensions of neurons, receive multiple spatio-temporal inputs from other neurons via synapses and play a vital role in single neuron computation. The integration of signals within dendrites enables diverse computations such as direction selectivity \cite{single1998dendritic,branco2010dendritic}, coincidence detection \cite{agmon1998role}, and logical operations \cite{koch2004biophysics,gidon2020dendritic}. The powerful computational abilities of individual biological neurons, particularly due to their dendritic arborizations, present a formidable challenge for the development of artificial neuronal models that can accurately capture the full range of dynamics exhibited by real neurons. 

Since the integration mechanisms of dendrites are complicated, the early models of neurons usually consider the point neuron which only contain the somatic structure but ignore the dendrites. For example, the McCoulloch and Pitts neuron model linearly sums all the synaptic inputs at the soma and generates the outputs through an activation function \cite{mcculloch1943logical}. This kind of artificial neuron is widely used in many common machine learning models, such as the Multilayer Perceptrons \cite{rosenblatt1958perceptron}, LeNet \cite{lecun1989backpropagation}, AlexNet \cite{krizhevsky2017imagenet} etc. However, there are both experimental and theoretical results indicating that dendrites process signals in a nonlinear manner and should be regarded as an independent computational unit \cite{hausser2000diversity,polsky2004computational,london2005dendritic,hao2009arithmetic, li2014bilinearity}. Hence the classical point neuron model is oversimplified to capture the full characteristics of the biological neurons and thus the feature of dendritic nonliearities should be taken into account.
Recently, some studies have attempted to incorporate dendrites into single neuron models \cite{poirazi2003pyramidal,gutig2006tempotron,memmesheimer2014learning,ujfalussy2018global,tzilivaki2019challenging,beniaguev2021single,beniaguev2022multiple}. However, there is still a lack of a simple and biologically interpretable model that can reflect both the sub- and supra-threshold behaviors of the biological neurons with dendritic structure.

In this paper, we introduce a novel two-layer neural network named as dendritic bilinear neural network (DBNN) that can accurately replicate the input-output(I/O) mapping of biological neurons. We train DBNN with the data from the biological neurons when receiving multiple spatio-temporal synaptic inputs. We show that DBNN can faithfully predict the somatic response including both the sub-threshold voltage and the spike time of the neuron. 95\% variance of the sub-threshold voltage can be explained and the precision of the spike time prediction can be greater than 80\%. Our DBNN is concise and the number of parameters is much less compared with multi-layer ANN. The predictive power and computational capacity of DBNN have been verified on different types of biological neuron models. Furthermore, we find that the trained parameters in DBNN are biologically interpretable which reflect both the single post-synaptic potential (PSP) response and a bilinear dendritic integraion rule for multiple synaptic inputs revealed in previous studies \cite{hao2009arithmetic,li2014bilinearity}. We can then use proper intial values of parameter based on the biological properties to accelerate the training speed. Moreover, we demonstrate that DBNN can characterize the dendritic computation power of biological neurons through solving direction selectivity and coincidence detection problems. And we also apply DBNN for image classification task and it can outperform the ANN with the same number of trainable parameters. Our work presents a comprehensive framework for how to incorporate dendritic features into a neural network.

\paragraph{Related works}
Previous works have attempted to map the dynamics of biological neurons with dendrites onto artificial neural networks (ANNs). While two-layer ANNs are successful in fitting the firing rate of hippocampus CA1 pyramidal neurons \cite{poirazi2003pyramidal} or fast spiking basket neurons \cite{tzilivaki2019challenging}, they cannot predict the sub-threshold voltage and exact spike time. To address these limitations, ANNs incorporating temporal convolutions have been developed for accurate predictions of spike time \cite{gutig2006tempotron,memmesheimer2014learning,beniaguev2022multiple}. Additionally, hierarchical cascade models have been proposed to capture the sub-threshold voltage of L2/3 pyramidal neurons \cite{ujfalussy2018global}. Recently, a state-of-the-art seven-layer temporal-convolutional network (TCN) \cite{bai2018empirical} has been developed that fully captures both sub-threshold voltage and spike time of L5 pyramidal neurons\cite{beniaguev2021single}, but at high computational cost due to the large number of parameters. Moreover, it is unclear how these parameters are related to the properties of biological neurons.

\section{Results}
To establish a mapping from the dynamics of a biological neuron to an ANN, we develop a  two-layer ANN called the dendritic bilinear neural network (DBNN) based on the features of single neuronal computation. DBNN is trained using the input-output data of a biological neuron model with dendritic structure simulated by NEURON software \cite{hines1997neuron} to fully capture both the sub-threshold voltage and spike time dynamics of the neuron.

\subsection{Dendritic bilinear neural network}

A single neuron receives thousands of synaptic inputs through its dendrites from other neurons. These inputs are integrated and transmitted to the soma, where output is generated. Given a total of N synapses on the dendrites receiving pre-synaptic spike trains $x_i(t)$, the somatic response $v(t)$ and spike time $\hat{t}$ can be theoretically calculated using Rall's cable theory \cite{rall1959branching, rall1964theoretical, rall1967distinguishing} by solving a large system of differential equations. However, this method requires expensive computational resources. Here, we aim to construct a biologically interpretable ANN that accurately captures the dynamics of biological neurons. Specifically, when presented with the same input $x_i(t), i=1,2,\cdots,N$ to a biological neuron, the proposed ANN will generate the output that precisely predicts both the somatic voltage and spike time of the biological neuron.

To develop an ANN that can accurately capture the intricate input-output relationships of biological neurons, we first describe the simplest case of how the neuronal voltage responds to a single synaptic input. To this end, we employ a double-exponential function (equation \eqref{biexp}) to describe the postsynaptic potential \cite{gutig2006tempotron, gerstner2014neuronal},
\begin{equation}\label{biexp}
    k_i(t)=\omega_i(1-e^{-t/\tau_{r,i}})e^{-t/\tau_{d,i}},
\end{equation}
where $i$ is the synapse index, $\omega_i$ is the weight of synaptic input and $\tau_{r,i}$ and $\tau_{d,i}$ are rising and decay time constant, respectively. We can then express the response at the synapse when multiple inputs are received at different times as the convolution of the corresponding pre-synaptic spike train $x_i(t)$ with the response kernel defined as $k_i(t), i.e.$,
\begin{equation}\label{Vi}
    v_i(t)=x_i(t)\otimes k_i(t).
\end{equation}

We now describe the scenario that a neuron receives inputs from multiple synapses located at different dendritic sites. Previous experimental results have suggested that a neuron integrates these inputs in a nonlinear manner \cite{poirazi2003arithmetic,polsky2004computational}. In contrast to studies that use  sigmoid functions to describe the integration process, here we attempt to characterize this process by utilizing the simplest possible nonlinear function: the quadratic polynomial. To be specific, given the voltage responses  $v_1(t), v_2(t),..., v_N(t)$ induced by individual input received at each single synapse, the integrated response  described by a quadratic integration function is as follows:
\begin{equation*}\label{bilinear}
    v(t)=\sum_{i=1}^Nv_i(t)+\sum_{j=1}^{N}\sum_{k=1}^{j-1}a_{jk}v_j(t)v_k(t)+v_0.
\end{equation*}
In equation \eqref{bilinear}, note that we exclude all the square terms $v^2_j$ and retain only the cross terms $v_jv_k$ in order to make the integration rule valid even when an individual input $i$ is given, i.e., the response is $v_i(t)$ in such a case. Based on equations \eqref{biexp}\eqref{Vi}\eqref{bilinear}, we build the two-layer dendritic bilinear neural network (DBNN) with its architecture illustrated in Figure 1(a). 

\subsection{Data generation and training protocal}
We first use DBNN to approximate the activity of a biological neuron. To generate the training data for DBNN, we simulate biologically realistic neuron models with dendrites using NEURON software \cite{hines1997neuron}. To be specific, we utilize three representative types of neuron models, namely a basal ganglion neuron \cite{sheasby1999impulse}, a layer 2/3 pyramidal neuron \cite{smith2013dendritic}, and a layer 5 pyramidal neuron \cite{hay2011models}. Each neuron is endowed with dendritic structure and active ion channels including $Na^+,K^+$, and $Ca^{2+}$. Excitatory and inhibitory synapses are randomly distributed on the dendrites in varying numbers depending on the specific neuron model, with $N=9, 749, 1278$ synapses for the basal ganglion, layer 2/3 pyramidal, and layer 5 pyramidal neurons, respectively. The excitatory synapses include both AMPA and NMDA receptors, and the inhibitory synapses include GABA-A receptors. Each synapse receives independent inhomogeneous Poisson spike trains. The input spike trains and the corresponding somatic voltage response of the neuron are recorded for subsequent use as input and output data to DBNN, respectively. Each stimulation lasts for six seconds with millisecond resolution, and the stimulation is repeated 1000 times for training and 100 times for testing with different initialization. The detailed information for the data generation can be found in the Supplementary Material.

In the training procedure of DBNN, we utilize the mean square error (MSE) as the loss function to quantify the difference between DBNN's output and the voltage of the biological neuron model. To optimize the parameters, the standard stochastic gradient descent (SGD) \cite{rumelhart1986learning} algorithm is employed. We set the learning rate to be $0.001$ and use a mini-batch size of 128. DBNN is trained for 1000 epochs. All of the parameters in DBNN, including the time constants $\tau_{r,i}$ and $\tau_{d,i}$, weight $\omega_i$, quadratic coefficient $a_{jk}$ and bias term $v_0$, are updated during training. The training process is carried out on an Nvidia A100 GPU and takes approximately 2-3 hours to complete.

\subsection{Predictive performance of DBNN}
Following the training process, we observe that DBNN is capable of accurately predicting the output of biological neurons, including the subthrehold voltage and spike time. To measure the performance of DBNN for predicting the sub-threshold voltage, we utilize the variance explained (VE) metric defined as
\begin{equation*}
    \text{VE}=1-\frac{\sum_i(y(t_i)-y'(t_i))^2}{\sum_i(y(t_i)-\mathbb{E}[y(t)])^2}
\end{equation*}
where $y(t_i)$ and $y'(t_i)$ are the true and predicted voltage for different discrete time and $\mathbb{E}(y(t))$ is the mean value of the true voltage. The closer the VE is to 1, the more accurate prediction made by DBNN is.

After training, DBNN is first applied to predict the sub-threshold voltage of a passive layer 2/3 pyramidal neuron without active ion channels, which yields the VE of approximately 99\%. For neurons with $Na^+,K^+$, and $Ca^{2+}$ channels included, the dendritic integration process becomes more nonlinear. In this case, the VE can still reach 95\% when AMPA receptors are used exclusively, and 93\% when both AMPA and NMDA receptors are used. (Figure 1(b)(c) and Table \ref{result compare}).

\begin{figure}[h]
    \centering
    \includegraphics[width=15cm]{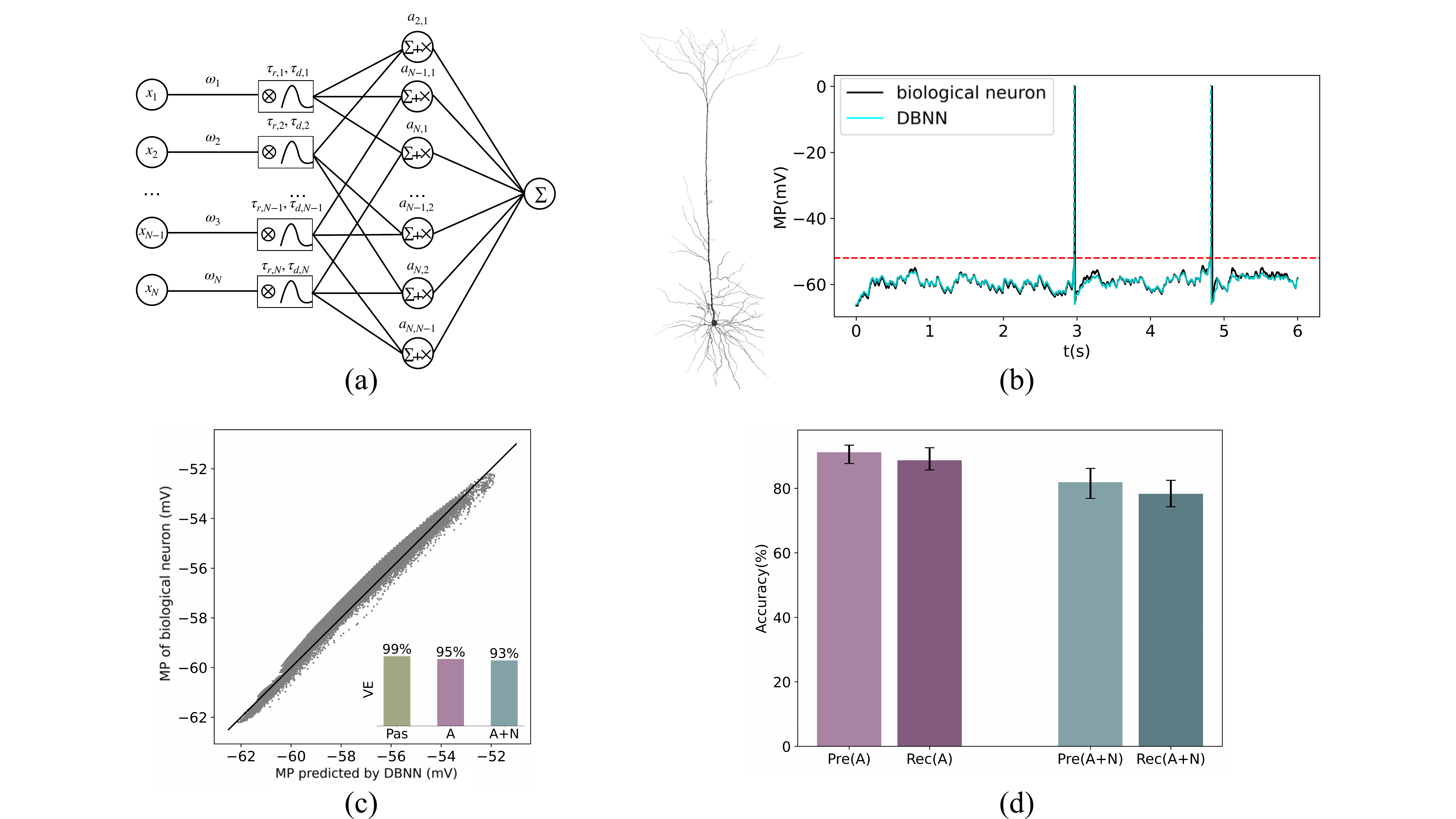}
    \caption{DBNN is capable of predicting the dynamics of layer 2/3 pyramidal neurons, including the sub-threshold voltage and the spike time. (a) The architecture of DBNN, where the symbol $\sum$ represents linear summation and $\sum+\times$ represents quadratic multiplication. The rest notations are the same as those described in equations \eqref{biexp}\eqref{Vi}\eqref{bilinear}. (b) Left: the morphology of the pyramidal neuron. Right: the output of DBNN (cyan) well agrees with the membrane potential (MP) of the pyramidal neuron model shown in Left (black). Red dashed line is the firing threshold. (c) The scatter plot of the predicted sub-threshold MP by DBNN versus that of the biological neuron model. Lower right is the bar plot of variance explained (VE) for passive neurons (Pas), active neurons with only AMPA receptors (A) and both AMPA and NMDA receptors (A+N). (d) The precision (Pre) and recall (Rec) values when using DBNN to predict the spike time of the neuron. Two different electrophysiology conditions are considered here as A and A+N in Figure 1(c).}
    \label{fig:1}
\end{figure}

We next use DBNN to predict spike time, which is a more challenging task because spike time results from a highly nonlinear process of action potential generation beyond quadratic nonlinearity in DBNN. To circumvent the difficulty of fitting action potentials, the prediction of spike time by DBNN is made when the subthreshold voltage crosses a firing threshold. To be specific, when the predicted output $v(t_n)$ at time $t_n$  satisfies that $v(t_n)\geq v_{th}$ and $v(t_{n-1})<v_{th}$, we then define $\hat{t}=t_n$ as the predicted spike time, where $t_{n-1}$ and $t_n$ are two consecutive discrete time points and $v_{th}$ is a firing threshold value. It shall  be noted that the DBNN does not account for a reset mechanism, which is another strong nonlinear effect in the biological neuron that the somatic voltage will rapidly drop after a spike due to the outflow of $K^+$. To further improve DBNN, we modify equation \eqref{bilinear} to incorporate a reset term in DBNN
\begin{equation}\label{reset}
    v(t)=\sum_{i=1}^Nv_i(t)+\sum_{j=1}^{N}\sum_{k=1}^{j-1}a_{jk}v_j(t)v_k(t)+v_0+v_{reset}\sum_l\Theta(t-\hat{t}_{l})e^{\frac{-(t-\hat{t}_l)}{\tau_{reset}}},
\end{equation}
where $v_{reset}$ is the amplitude of voltage reset after a spike, $l$ is the index of spike, $\Theta$ is the Heaviside function and $\tau_{reset}$ is the reset time constant. 

The accuracy of spike time prediction is then evaluated by using the metrics of precision and recall. We treat a predicted spike time $\hat{t}$ as a true positive (TP) case if the true spike time is within a 10 millisecond time window of $\hat{t}$. Then precision is calculated by the ratio of the number of true positives to the sum of true positives and false positives, while recall is calculated by the ratio of the number of true positives to the sum of true positives and false negatives. For the case that neurons with only AMPA receptors, a precision rate of $91\%$ and a recall rate of 89\% can be achieved, while for the biological neuron model equipped with both AMPA and NMDA receptors, the precision rate is 82\% and the recall rate is 78\%. These metrics suggest that DBNN can make predictions for spike time with high accuracy
(Figure 1(d) and Table \ref{result compare}).

To show the advantage of DBNN, we next perform a comparative analysis of DBNN and other ANNs developed recently \cite{ujfalussy2018global,beniaguev2021single} in predicting a biological neuron's activity. We compare these models in various cases, including passive membranes, active membranes only with AMPA receptors (A), and with both AMPA and NMDA receptors (A+N). The results, summerized in Table \ref{result compare}, demonstrate that DBNN outperforms the hierarchical cascade of linear-nonlinear model (hLN) \cite{ujfalussy2018global} in predicting sub-threshold voltage and the hLN fails to predict spike time. When compared to the state-of-the-art seven-layer TCN \cite{beniaguev2021single}, DBNN can achieve similar prediction accuracy for both sub-threshold voltage and spike time but with much fewer parameters ($\mathcal{O}(10^5)$ in DBNN versus $\mathcal{O}(10^7)$ in seven-layer TCN). The results suggest that DBNN provides a simple and effective framework to accurately capture the dynamics of a biological neuron with dendritic structures.

\begin{table}[H]
  \caption{Results compared with related works}
  \label{result compare}
  \centering
  \begin{tabular}{ccccc}
    \toprule
    Neuron type     & Objective     & DBNN & hLN\cite{ujfalussy2018global} & TCN\cite{beniaguev2021single} \\
    \midrule
    Passive & Sub-threshold voltage & \textbf{VE:0.99} & VE:0.95 & \textbf{VE:0.99} \\
    \midrule
    \multirow{3}*{Active(A)} & Sub-threshold voltage  & \textbf{VE:0.95}  & VE:0.92 & \textbf{VE:0.95}   \\
    ~     & \multirow{2}*{Spike time} & \textbf{Pre:0.91}    & \multirow{2}*{N/A} & Pre:0.64  \\
    ~     & ~       & \textbf{Rec:0.89} & ~ & Rec:0.64 \\
    \midrule
    \multirow{3}*{Active(A+N)} & Sub-threshold voltage  & VE:0.93  & VE:0.91 & \textbf{VE:0.94}   \\
    ~     & \multirow{2}*{Spike time} & \textbf{Pre:0.82}    & \multirow{2}*{N/A} & Pre:0.58  \\
    ~     & ~       & \textbf{Rec:0.78} & ~ & Rec:0.60 \\
    \midrule
    \multirow{2}*{Active(A+N)} & Number of parameters  & $\mathcal{O}(10^5)$  & $\boldsymbol{\mathcal{O}(10^3)}$ & $\mathcal{O}(10^7)$   \\
    ~     & Running time for training & \textbf{2-3 hours}    & \textbf{2-3 hours} & 2-14 days  \\
    \bottomrule
    \multicolumn{5}{c}{Note: N/A means that the model cannot predict the spike time}
  \end{tabular}
\end{table}

The DBNN also possesses a variety of generalization ability. After training on a dataset in which all synapses are activated in the biological neuron model, DBNN can successfully predict the somatic sub-threshold voltage for entirely different input patterns, such as varied synaptic input frequency or the activation of partial synapses (Figure S1). Furthermore, the DBNN's predictive capacity is verified across different biological neuron types, including ganglion and pyramidal neurons (Figure S1).

\section{Biological interpretation of DBNN}
We next demonstrate another advantage of DBNN that the parameters after training are biologically interpretable. The parameters capture the key features of the post-synaptic potential (PSP) induced by a single input and a bilinear dendritic integration rule of synaptic inputs - both are important characteristics of biological neurons. The biological interpretability of DBNN suggests that DBNN can effectively exploit dendritic features to achieve computational capabilities.

\subsection{DBNN captures the post-synaptic potentials}
\label{single}
To understand how DBNN can successfully predict the activity of a biological neuron, we first examine the input kernel $k_i(t)$ (as equation \eqref{biexp}) in DBNN after training. This kernel reflects the output when only an input spike $x_i(t)$ is given while no input is given to all other nodes. Therefore, it may relate to the postsynaptic potential of a biological neuron that receives only an individual synaptic input. To investigate this relation, we simulate each synapse located at different positions of the dendritic branches and measure the corresponding PSP at the soma of the biological neuron. Subsequently, we compare the PSPs with the double exponential kernels based on the parameters $\omega_i$, $\tau_{r,i}$, and $\tau_{d,i}$ after training using the data from the same neuron. Interestingly, our results reveal a significant similarity between the PSPs and the double exponential kernels, which is not trivial considering that the training dataset only involves the scenario that all synapses are activated, but not include the scenario of a single input is given. Furthermore, the similarity holds for both the activation of an individual excitatory synapse and inhibitory synapse (Figure 2(a) and Figure 2(b)).

These findings suggest that using the double-exponential form as the input kernel is valid, given its ability to capture how an individual synaptic input propagates to the soma. Furthermore, our results indicate that the parameters in the double-exponential kernel can be trained effectively, even in the case that a single input is not presented in the training data. Our DBNN's training setup associates each index $i$ with a specific synaptic location on the dendrites. In turn, the weight $\omega_i$ represents the PSP magnitude activated by this corresponding synapse, while the time constants $\tau_{r,i}$ and $\tau_{d,i}$ are related to the location of the dendrites where the synapse is located. In general, the farther a synapse is located from the soma, the larger its associated time constant will be. Therefore, the parameters of the double-exponential kernel can reflect the synaptic input location and the characteristics of single PSPs of the biological neuron.

\subsection{DBNN captures the dendritic integration rule}
\label{multiple}
As the input kernel corresponds to a single PSP of a biological neuron, the next layer in DBNN described as the quadratic function in equation\eqref{bilinear} is supposed to relate to the non-linear integration of multiple synaptic inputs. Previous investigations \cite{hao2009arithmetic,li2014bilinearity}, which entailed electrophysiological experiments and theoretical analyses, reveal that the integration of a pair of synaptic inputs follows a bilinear rule which is in the form of
\begin{equation}
\begin{aligned}
\label{Li}
    V_S(t) &\approx V_1(t)+V_2(t)+\kappa(t)V_1(t)V_2(t),
\end{aligned}
\end{equation}
where $V_1(t)$ and $V_2(t)$ are the PSPs when the two synaspes are activated individually, and $V_S(t)$ is the somatic response to the same pair of synaptic inputs activated simultaneously, $\kappa(t)$ is the integration coefficient for the pair of synaptic inputs that only depends on the location of the inputs but not their strengths. And this rule is further generalized to all types of inputs, including a pair of excitatory inputs, a pair of inhibitory inputs, and multiple excitatory and inhibitory inputs. In the generalized case, the integration coefficients are denoted as $\kappa_{EE}(t)$, $\kappa_{EI}(t)$, and $\kappa_{II}(t)$.

Upon observing the similarities between the quadratic function designed in DBNN (equation \eqref{bilinear}) and the bilinear dendritic integration rule (equation \eqref{Li}), we began to contemplate whether DBNN could be trained to learn the bilinear dendritic integration rule. We first measure the values of the integration coefficients ($\kappa_{EE}(t)$, $\kappa_{EI}(t)$, and $\kappa_{II}(t)$) in the biological neuron by giving a pair of inputs first separately and then simultaneously to all possible pairs of synapses. Then we compare these integration coefficients with the corresponding quadratic coefficients $a_{ij}$ in DBNN (Equation\ref{bilinear}) after training. We find that the quadratic coefficients well matches with the integration coefficients (Figure 2(c) and Figure 2(d)). This finding suggests that DBNN accurately employs the bilinear dendritic integration rule for the integration of multiple synaptic inputs.

From the above, we have gained insight into how the dynamics of the real biological neuron with dendrites can be captured by DBNN. For a single synaptic input at different locations, the double-exponential kernels in DBNN learn the shape of both excitatory and inhibitory PSPs of the biological neuron. For multiple inputs from varies locations, the DBNN learns the bilinear rule to integrate all PSPs in a simple nonlinear manner. As a result, the sub-threshold voltage is well-fitted by the DBNN, leading to the moderately accurate inference that each time the membrane potential reaches the firing threshold, the neuron will emit a spike. The result highlights the biological interpretability of DBNN.

\begin{figure}[h]
    \centering
    \includegraphics[width=15cm]{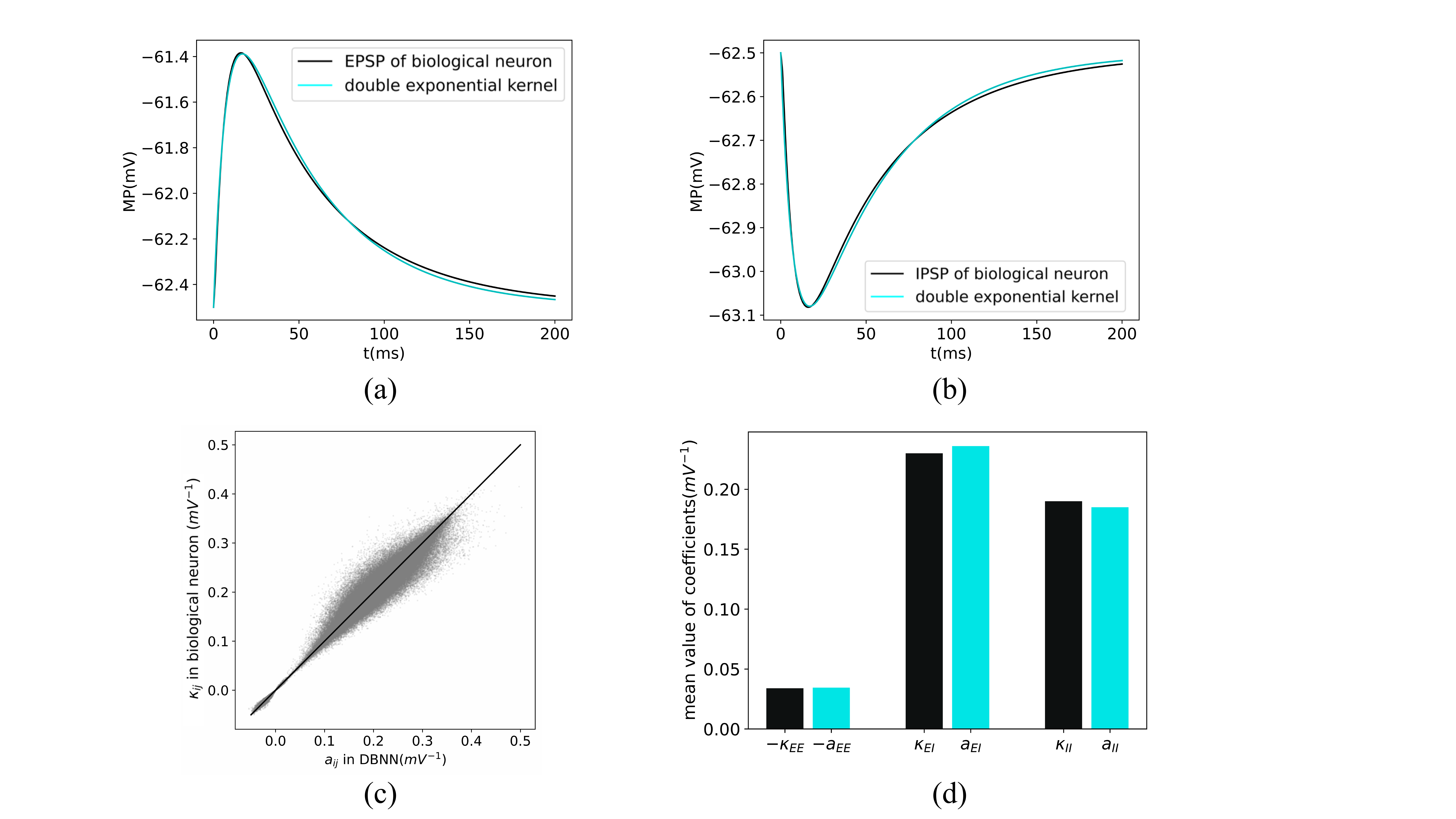}
    \caption{Biological interpretation of the parameters in DBNN after training. (a) The well match between the PSP of the biological neuron induced by a single excitatory synaptic (black) input and the double exponential kernel in DBNN (cyan). (b) As in Figure 2(a), the double exponential kernel (cyan) also matches the corresponding inhibitory PSP (IPSP) curve (black). (c) The scatter plot of the bilinear integration coefficient $\kappa_{ij}$ for each pair of synaptic locations w.r.t. the corresponding quadratic coefficient $a_{ij}$ in DBNN. (d) The mean values of the bilinear integration coefficient and the corresponding quadratic coefficient $a_{ij}$ for different pairs of excitatory (E) or inhibitory (I) synapse including EE, EI or II.}
    \label{fig:my_label}
\end{figure}

\subsection{Training speed and low dimensional structure}
The number of parameters in DBNN is much less in comparison to deep neural networks with many layers. However, owing to the quadratic function specified in equation \eqref{bilinear}, the number of parameters contained in DBNN exceeds that of the linear integration or sigmoid nonlinearity as utilized in Ujfalussy et al. \cite{ujfalussy2018global}. Without additional constraints, training DBNN requires more computational time. Nonetheless, Sections \ref{single} and \ref{multiple} have uncovered that parameters of DBNN are biologically interpretable, which inspires one to implement appropriate strategies that expedite the training process. For example, all the rising time constant $\tau_{r,i}$ for different kernels can be initialized to $5ms$, an biologically meaningful value for synaptic time constants. By adopting this method of initialization, the training time of DBNN will be reduced considerably and even be closed to models employing sigmoid nonlinearity (Table \ref{result compare} and Figure S2).

Furthermore, Beniaguev et al. \cite{beniaguev2022multiple} have discovered that the PSPs induced by synaptic inputs at various sites of the biological neuron exhibit low dimensional structures that can be reduced to 3-dimensional vectors. Consequently, if we use the 3-dimensional vectors to represent all the PSPs of the real neuron, we can transform the original problem into an much easier linear regression problem. This approach also facilitates acceleration of the training speed (Figure S2). The proof of turning into a linear regression problem is illustrated in the Supplementary material.

\section{Applications}
Biological neurons with dendritic morphology are capable of performing the computation of direction selectivity and coincidence detection. Here, we also show that our DBNN is able to perform these tasks. We further utilize DBNN to address the MNIST classification task and its can outperform the traditional two-layer ANN. These results indicate that DBNN is capable of capturing the complex dendritic computation power of biological neurons.

\subsection{Direction selectivity}
\label{para_direc}
According to the experiments by Branco et al. \cite{branco2010dendritic}, dendrites of cortical pyramidal neurons exhibit selectivity for the direction of spatio-temporal synaptic inputs. Specifically, if the a sequence of excitatory synaptic inputs are received from the distal dendrite toward the soma which is the preferred direction, they are more likely to promote somatic firing. Conversely, if inputs are received in non-preferred direction, they will less likely to generate a somatic spike (Figure 3(a)).

Notably, direction selectivity can also be present in DBNN. After training to fit the response of a biological neuron, we perform the following direction selectivity experiment: stimulating the input nodes of DBNN associated with different synapses leads to emit a spike when the synaptic inputs are activated in the preferred direction. However, when the synapses are activated in the opposite (i.e., non-preferred) direction, the resulting output of DBNN is insufficient to exceed the threshold required for spike generation (Figure 3(a)).

\begin{figure}[ht]
\centering
\subfigure[]{
\includegraphics[width=5.5cm]{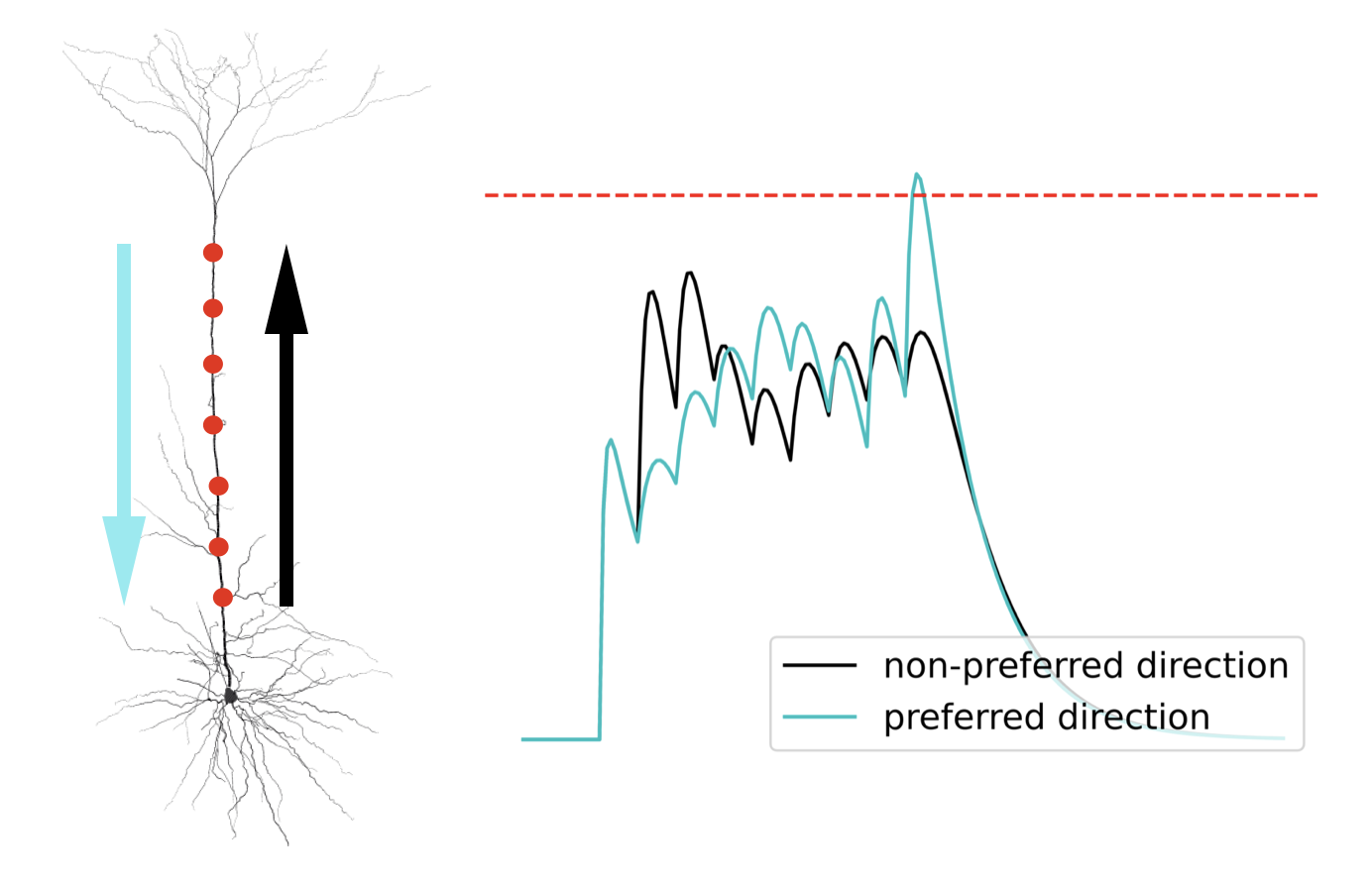}
\label{direction}
}
\quad
\subfigure[]{
\includegraphics[width=7cm]{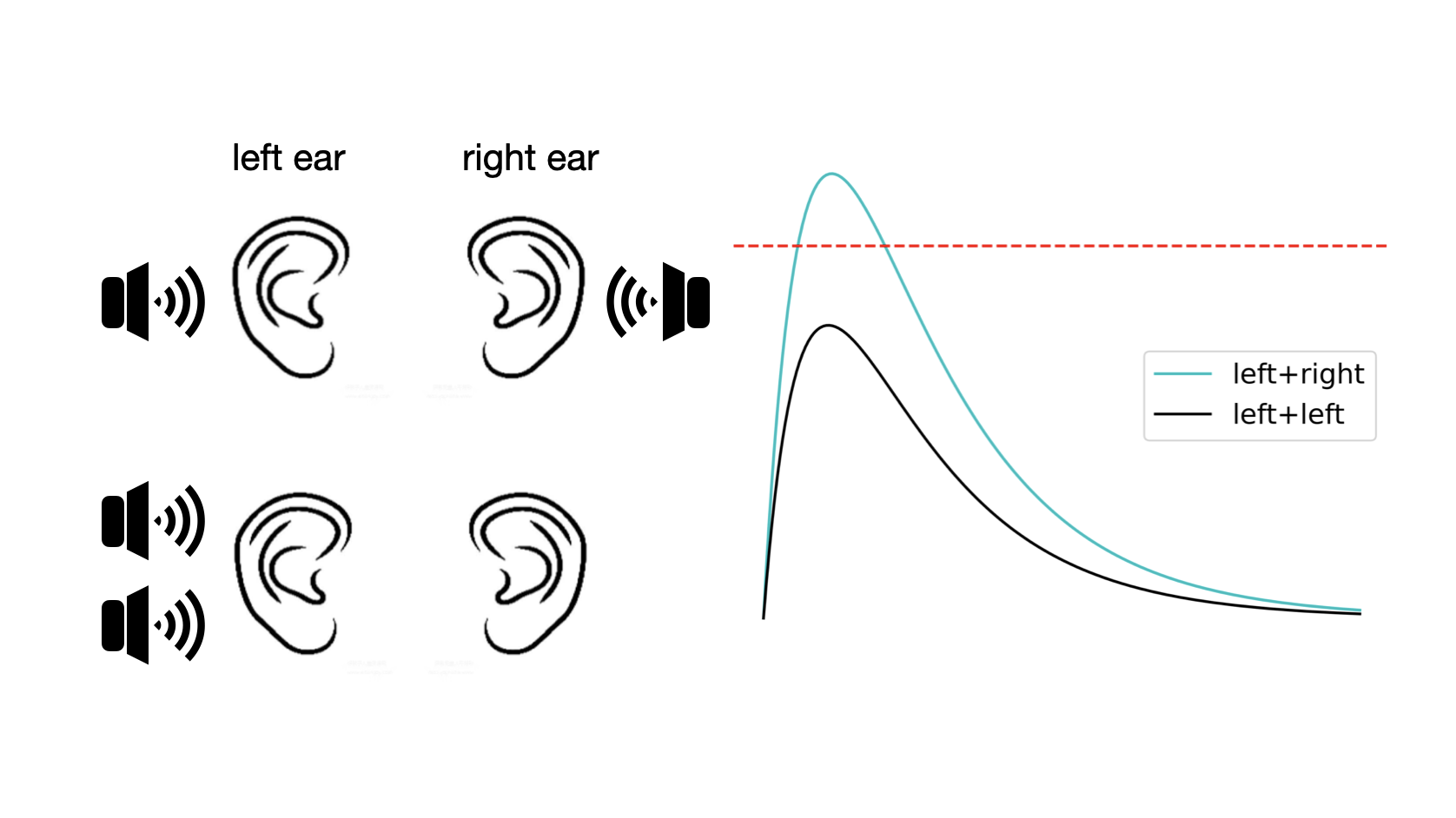}
\label{coincidence}
}
\quad
\subfigure[]{
\includegraphics[width=3.2cm]{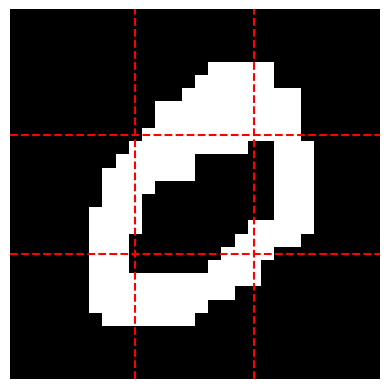}
\label{segment}
}
\quad
\subfigure[]{
\includegraphics[width=3cm]{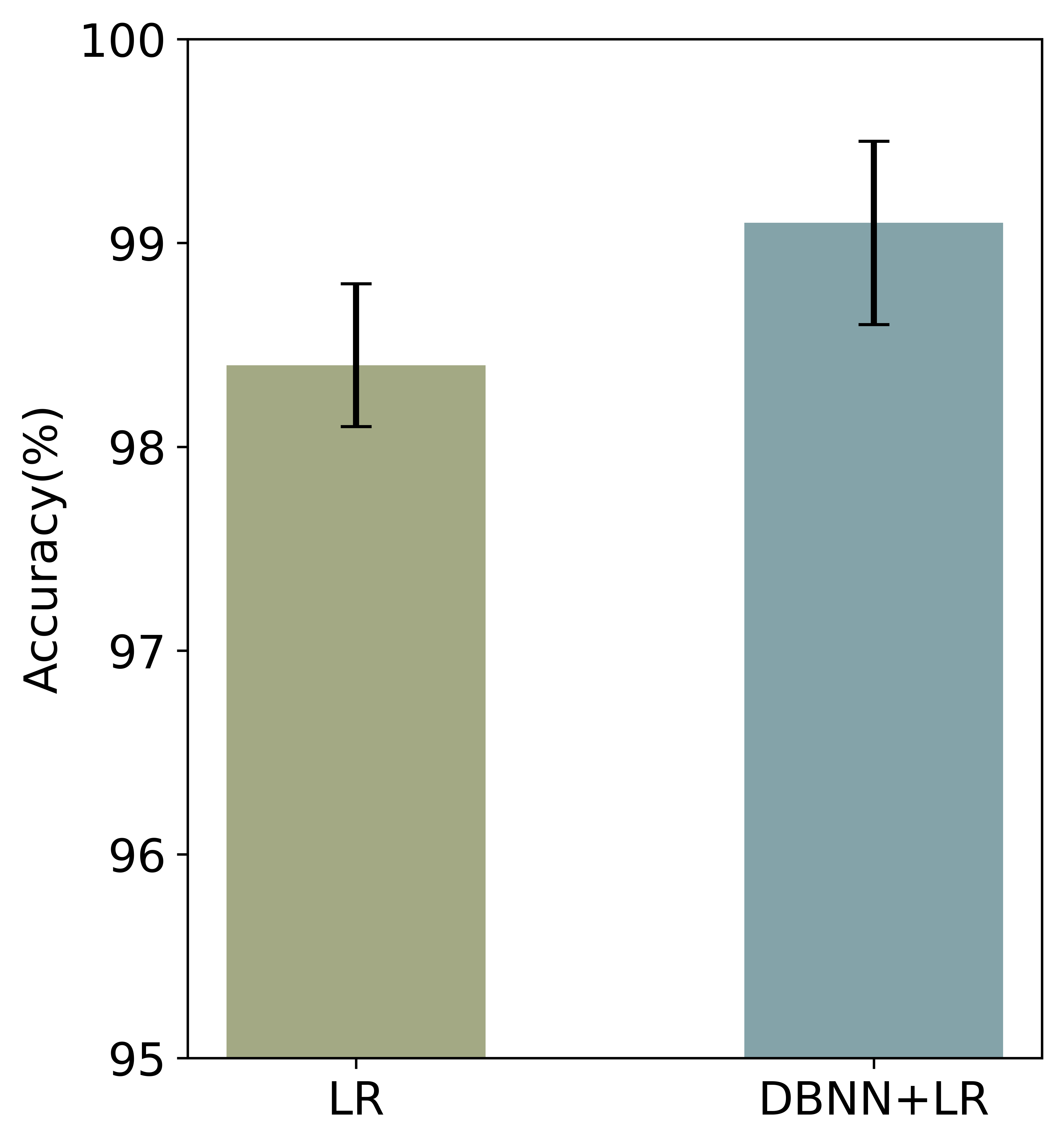}
\label{binary}
}
\quad
\subfigure[]{
\includegraphics[width=4cm]{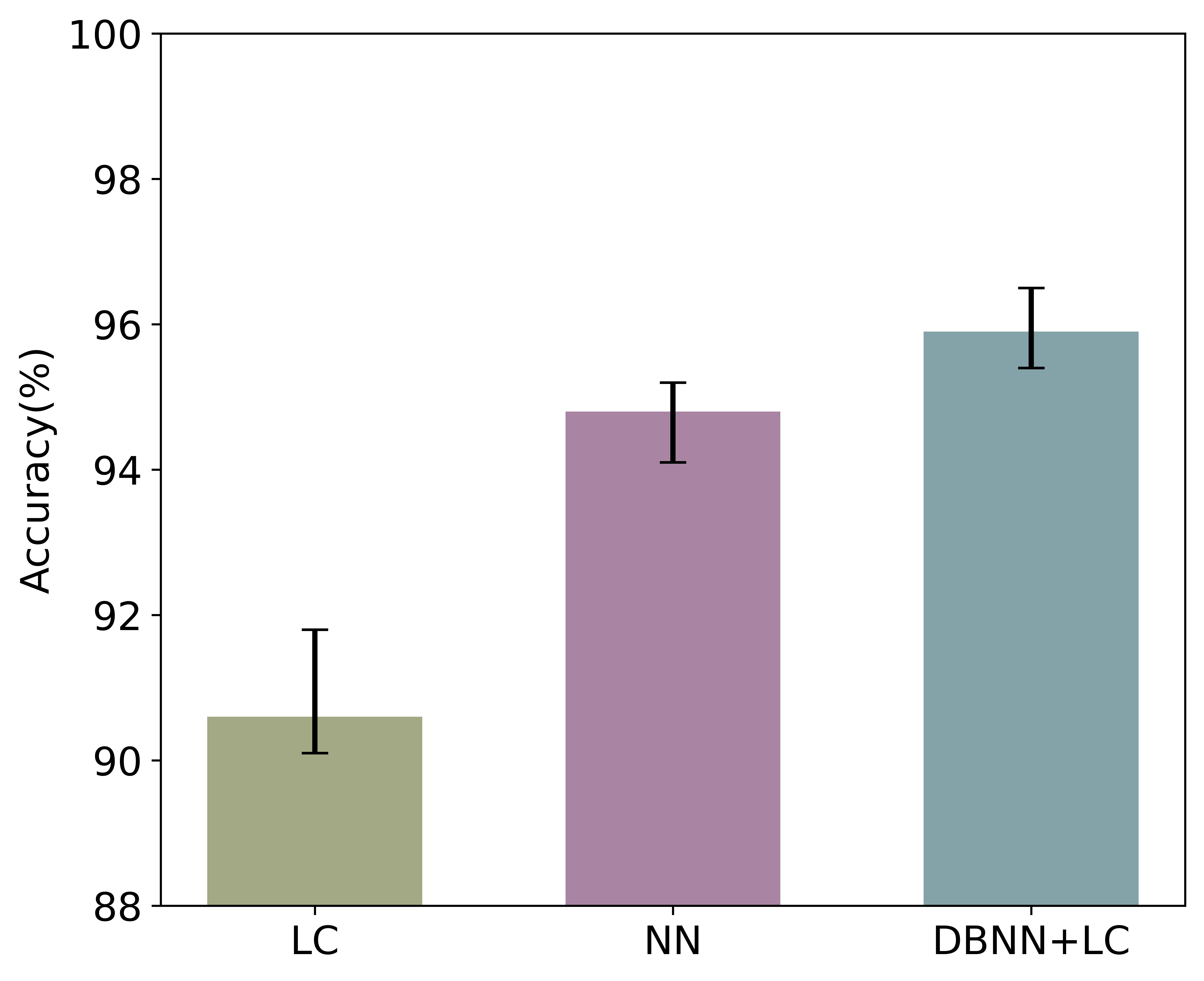}
\label{ten}
}
\caption{Applications of DBNN for direction selectivity and coincidence detection and MNIST classification. (a) Left: the illustration for direction selectivity. Right: the realization of direction selectivity in DBNN with the synaptic inputs for the preferred direction (cyan) and the non-preferred direction (black). Red dashed line is the firing threshold. (b) Left: the illustration for coincidence detection. Right: the realization of coincidence detection in DBNN with the synaptic inputs for the different sides (cyan) and the same side (black). (c) Divide the original MNIST digit into nine equal parts. (d) DBNN with logistic regression (LR) outperforms using LR directly when solving MNIST binary classification task. (e) DBNN with linear classifier (LC) outperforms using LC directly or two-layer neural network(NN) when solving MNIST 10-category classification task. }
\end{figure}

\subsection{Coincidence detection}
In addition to direction selectivity, biological neurons with dendrites can also conduct the coincidence detection. For example, in the early auditory pathway, sound detected from the left and right ear generates synaptic inputs located on different dendritic branches of bipolar neurons \cite{agmon1998role}. These neurons can then detect the input coincidence that whether the sounds detected come from different sides (Figure 3b).

We find that DBNN is also capable of performing coincidence detection. As the same with Section \ref{para_direc}, the well-trained DBNN is used to verify this capability. When two synaptic inputs are simultaneously activated at separate dendritic branches, the resulting DBNN's response is stronger enough to generate a spike. In contrast, if we activate two synaptic inputs located at the same dendritic branch, the DBNN's response can not reach the firing threshold (Figure 3b). In this way, DBNN can perform the coincidence detection. Our findings indicate that DBNN exhibits dendritic computational power that is similar to the biological neurons.

\subsection{MNIST classification}
The MNIST classification task \cite{lecun1998gradient} is a well-known benchmark problem in the field of machine learning. Each sample in this dataset represents a $28\times 28$ pixels image that corresponds to a handwritten digit ranging from 0 to 9. We transform the original MNIST dataset so that it can be used as input to DBNN. To achieve this, we first remove the first row and column from each image, resulting in $27\times 27$ pixels images. Next, we convert the original gray-scale map to a black-white map by setting all pixels with values greater than 0 to 1 and segment the images into 9 regions (as shown in Figure 3(c)). We then flatten each region of $9\times9$ pixels into a row vector and expand it by making 9 consecutive copies of each element. In this way, each handwritten digit is associated with a $9\times 729$ spatio-temporal sequence, which can be served as the input for DBNN through 9 distinct synapses.

Initially, we train DBNN on the dataset generated from the ganglion neurons that had 9 distinct synapses on the dendrites \cite{sheasby1999impulse}. Once DBNN's parameters are fixed, we give the modified spatio-temporal sequences from MNIST dataset as input into DBNN and we collect the generated spike train corresponding to each digit. Consequently, we obtain various spike trains with 729 milliseconds duration that served as decoding signal for the original MNIST digits. Here, we consider both a simpler binary classification task and the traditional ten-category classification task. For the former problem, we utilize logistic regression to classify the spike trains produced by DBNN and compare with the results obtained from directly using logistic regression on the MNIST digits. Notably, the classification accuracy can be improved to $99\%$ (as illustrated in Figure 3(d)). As for the ten-category classification task, we employ a linear classifier to process the spike trains generated by DBNN and make comparisons with the results obtained solely via the linear classifier or a two-layer ANN with 10 hidden neurons (to maintain the number of trainable parameters the same) and relu activation function. Again, the using of DBNN results in a higher accuracy (Figure 3(e)). The detailed information for the logistic regression, linear classifier and two-layer ANN can be found in the Supplementary Material.

\section{Discussion}
We have presented a dendritic bilinear neural network (DBNN) that can accurately capture the somatic dynamics of biological neurons with dendritic structures, including sub-threshold voltage and spike time. We further demonstrate that DBNN's parameters are biologically interpretable. The double exponential kernels in the first layer of DBNN represent post-synaptic potentials for single synaptic input located at different dendritic branches. Meanwhile, the second layer's quadratic terms reflect the dendritic bilinear integration rule for multiple synaptic inputs. Additionally, we find that DBNN has the capacity to perform direction selectivity and coincidence detection which demonstrates that DBNN's dendritic computation power can approach the level of biological neurons. We also utilize DBNN to solve the image classification task, outperforming the traditional two-layer ANN.

Although we have shown that DBNN can capture the dynamics of different biological neuron types including ganglion neurons and pyramidal neurons, there are other types of neurons with diverse morphologies and electrophysiology properties. It remains an open question whether DBNN can capture the dynamics of all types of neurons or we should modify its structure to apply to different biological neuron models. And the integration of DBNN into existing deep learning frameworks as computational units is also an interesting direction which could possibly enhance the computational capacity of the ANN models. Our work provides a method of mapping dendritic function into a biologically interpretable neural network and the neural network can gain more computational power through this way. 

\section*{Acknowledgments}
This work was supported by Science and Technology Innovation 2030 - Brain Science and Brain-Inspired Intelligence Project with Grant No. 2021ZD0200204 and the Lingang Laboratory Grant No. LG-QS-202202-01 (S.L., D.Z.,); National Natural Science Foundation of China Grant 12271361 (S.L.); National Natural Science Foundation of China with Grant No. 12071287, 12225109 (D.Z.), Shanghai Municipal Science and Technology Major Project 2021SHZDZX0102 and the Student Innovation Center at Shanghai Jiao Tong University (S.L., D.Z.).

{
\small
\setcitestyle{numbers} 
\bibliographystyle{unsrt} 
\bibliography{ref.bib} 
}


\end{document}